%% file: main.tex
\documentclass[sigconf]{acmart}
\AtBeginDocument{%
  \providecommand\BibTeX{{%
    \normalfont B\kern-0.5em{\scshape i\kern-0.25em b}\kern-0.8em\TeX}}}

\copyrightyear{2021}
\acmYear{2021}
\setcopyright{acmlicensed}\acmConference[CHI '21]{CHI Conference on Human Factors in Computing Systems}{May 8--13, 2021}{Yokohama, Japan}
\acmBooktitle{CHI Conference on Human Factors in Computing Systems (CHI '21), May 8--13, 2021, Yokohama, Japan}
\acmPrice{15.00}
\acmDOI{10.1145/3411764.3445507}
\acmISBN{978-1-4503-8096-6/21/05}

\usepackage{booktabs}
\usepackage{tabularx}

\newcommand{\heffect}{\textsc{\textbf{H-Main}}}
\newcommand{\hcrowd}{\textsc{\textbf{H-Support 1}}}
\newcommand{\hsenti}{\textsc{\textbf{H-Support 2}}}
\newcommand{\haccept}{\textsc{\textbf{H-Support 3}}}
\def\cs{\textbf{S}}
\def\cq{\textbf{Q}}
\def\cc{\textbf{Q+S}}
\def\sq{\textbf{S\textrightarrow Q}}
\def\qs{\textbf{Q\textrightarrow S}}

\definecolor{OkabeItoBlue}{HTML}{2a6aa8}
\definecolor{OkabeItoPink}{HTML}{a84c82}
\definecolor{mygreen}{HTML}{00a85e}

\newcommand{\new}[2][1]{#2}

\begin{document}

\title[Ask Me or Tell Me? Enhancing the Effectiveness of Crowdsourced Design Feedback]{Ask Me or Tell Me? Enhancing the Effectiveness of Crowdsourced Design Feedback}

\author{Fritz Lekschas}
\email{lekschas@seas.harvard.edu}
\orcid{1234-5678-9012}
\affiliation{%
  \institution{Harvard School of Engineering and Applied Sciences}
  \city{Cambridge}
  \state{MA}
  \country{USA}
}

\author{Spyridon Ampanavos}
\email{sampanavos@gsd.harvard.edu}
\affiliation{%
  \institution{Harvard Graduate School of Design}
  \city{Cambridge}
  \state{MA}
  \country{USA}
}

\author{Pao Siangliulue}
\email{pao@b12.io}
\affiliation{%
  \institution{B12}
  \city{New York City}
  \state{NY}
  \country{USA}
}

\author{Hanspeter Pfister}
\email{pfister@seas.harvard.edu}
\affiliation{%
  \institution{Harvard School of Engineering and Applied Sciences}
  \city{Cambridge}
  \state{MA}
  \country{USA}
}

\author{Krzysztof Z. Gajos}
\email{kgajos@eecs.harvard.edu}
\affiliation{%
  \institution{Harvard School of Engineering and Applied Sciences}
  \city{Cambridge}
  \state{MA}
  \country{USA}
}

\renewcommand{\shortauthors}{Lekschas, et al.}

\begin{abstract}
  Crowdsourced design feedback systems are emerging resources for getting large amounts of feedback in a short period of time.
  Traditionally, the feedback comes in the form of a declarative statement, which often contains positive or negative sentiment.
  Prior research has shown that overly negative or positive sentiment can strongly influence the perceived usefulness and acceptance of feedback and, subsequently, lead to ineffective design revisions.
  To enhance the effectiveness of crowdsourced design feedback, we investigate a new approach \new{for mitigating the effects of negative or positive feedback by} combining open-ended and thought-provoking questions with declarative feedback statements. We conducted two user studies to assess the effects of question-based feedback on the sentiment and quality of design revisions in the context of graphic design.
  We found that crowdsourced question-based feedback contains more neutral sentiment than statement-based feedback. \new{Moreover, we provide evidence that} presenting feedback as questions followed by statements leads to better design revisions than question- or statement-based feedback alone.
\end{abstract}




\maketitle

\input{main-1-introduction-final}
\input{main-2-related-work-final}
\input{main-3-approach-hypotheses-final}
\input{main-4-manipulation-check-final}
\input{main-5-user-study-final}
\input{main-6-discussion-final}
\input{main-7-conclusion-final}

\begin{acks}
We would like to express our gratitude to Humphrey Obuobi for his help with the pilot study. Also, we thank all the participants who took part in our user studies. This research was supported in part by a gift from Adobe Research. The second author is partially funded by an Onassis Scholarship (Scholarship ID: F ZO 002/1 -- 2018/2019).
\end{acks}

\bibliographystyle{ACM-Reference-Format}
\bibliography{refs}

\end{document}

%% file: main-1-introduction-final.tex
\section{Introduction}

Feedback is a central part of learning and achievement that can help evaluate one's work, uncover problems, and promote new ideas for improvement. Yet, its effectiveness greatly varies by type and how it is framed, and its impact can be either positive or negative~\cite{hattie2007power}. In graphic design, \new{feedback is a vital part of the iterative design process and is typically solicited in critique sessions. However, these sessions are time and resource intensive.} Moreover, feedback \new{from alternative sources like} peers and online communities can be scarce~\cite{marlow2014rookie,xu2014voyant,luther2015structuring}, biased~\cite{tohidi2006getting,xu2014voyant}, and superficial~\cite{willett2012strategies,xu2012you}. Crowdsourced online feedback is an emerging mechanism to gather large amounts of feedback quickly~\cite{luther2014crowdcrit,greenberg2015critiki,yen2016social}. When structured appropriately, crowdsourced feedback can be as effective as expert feedback~\cite{yuan2016almost} and help designers produce more and better design revisions than they could have done otherwise~\cite{xu2014voyant,xu2015classroom,luther2015structuring}.

For crowdsourced feedback to be effective, it needs to foster productive reflection on the design to generate useful ideas for design revisions. Furthermore, the feedback needs to be acceptable to the designer, or else they will ignore it.
However, this is challenging because there is a tension between the productive value of feedback and acceptability, which is related to the feedback's perceived sentiment. For instance, Crain et al.~\cite{crain2017share} found that feedback with positive sentiment, which we will refer to as positive feedback, is typically preferred by content creators. However, positive feedback is less likely to lead to improvements through iteration. On the other hand, in their study, feedback with negative sentiment encouraged more design iterations but tended to have lower acceptance.
In the worst case, feedback with negative sentiment, which we will refer to as negative feedback, influences the recipient's affective state~\cite{baumeister2001bad,wu2018soften} and can reduce their overall task performance~\cite{cairns2014influence}. 

\begin{figure*}
  \centering
  \includegraphics[width=0.95\textwidth]{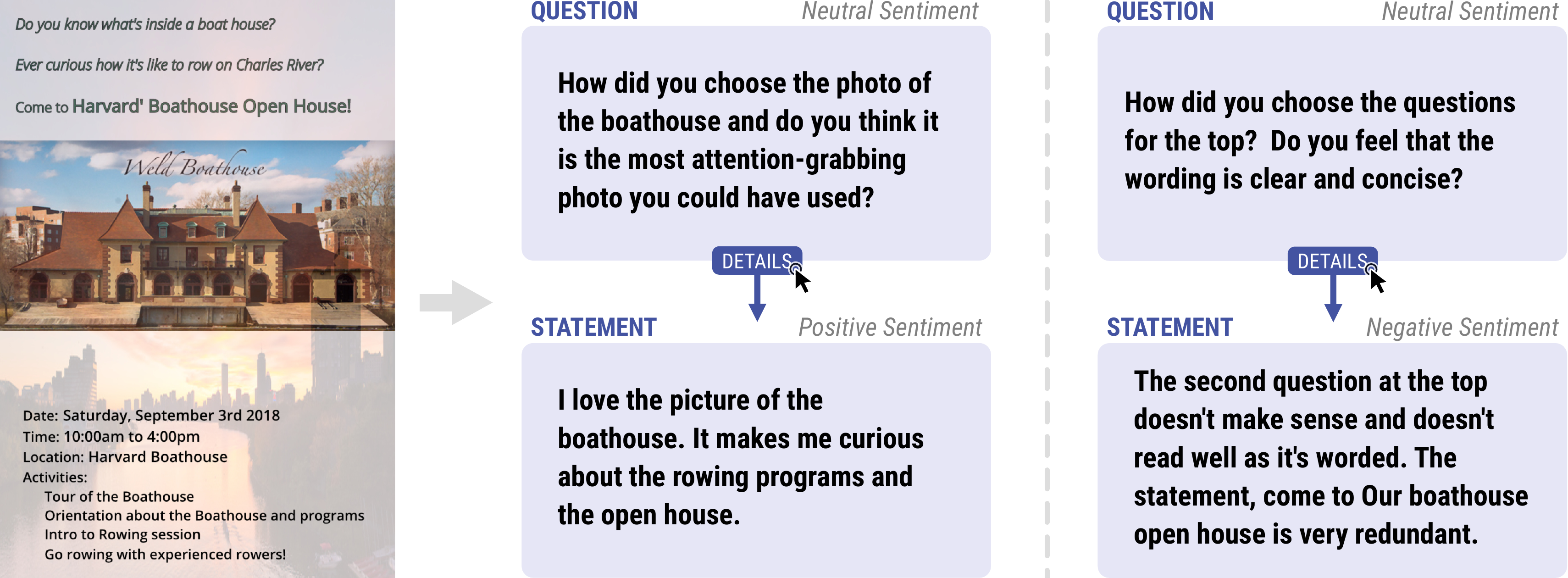}
  \Description[Two example feedback items for a flyer from the first user study]{Each feedback item consists of an open-ended question followed by a traditional statement. Although the related questions and statements target the same aspects of the flyer design, the questions carry more neutral sentiment than the statements.}
  \caption{\textbf{Enhanced Design Feedback:} Two example feedback items for a flyer from the first user study (\autoref{sec:manipulation-check}). Each feedback item consists of an open-ended question followed by a traditional statement. Although the related questions and statements target the same aspects of the flyer design, the questions carry more neutral sentiment than the statements.}
  \label{fig:example-feedback}
\end{figure*}

To improve the effectiveness of crowdsourced feedback on design revisions, we contribute a novel approach of enhancing traditional statement-based feedback with open-ended and thought-provoking questions (\autoref{fig:example-feedback}).
We hypothesized that presenting feedback in the form of a question followed by a statement would result in higher-quality design revisions compared to statement-based or question-based feedback alone. Building on prior work from several fields, our rationale for this hypothesis is twofold. First, we hypothesized that feedback in the form of open-ended questions carries less sentiment than statements and, subsequently, improves the acceptance of the feedback. Second, we hypothesized that the preceding open-ended question promotes productive reflection even if the statement-based feedback is superficial or unacceptable to the designer.

In design, reflection is fundamental in evaluating the current state of one's work relative to its goals and for generating ideas for improvements~\cite{schon1984reflective}. 
It is suggested that combining feedback with reflection is a superior format~\cite{brandt2008integrating} compared to feedback alone. For instance, feedback that incorporates a reflective task can lead to more extensive revisions and increased quality~\cite{yen2017listen} compared to traditional feedback.
An effective way to promote reflection is facilitative questioning.
For example, in teaching, questioning is known as an effective technique to trigger reflection and critical thinking among students~\cite{carnine1982effects,tofade2013best}.
However, questioning should not be the only type of feedback as it can otherwise irritate students~\cite{berghmans2012facilitative}.
Besides reflection, questions could balance the acceptance of feedback statements, assuming they contain neutral sentiment. \new{For instance, ordering feedback from positive to negative} has been shown to lead to a more balanced perception of negative feedback by improving the recipients' happiness and excitement~\cite{wu2017bitter}.

We conducted two online user studies in the context of graphic design to study the effects of enhancing statement-based with question-based feedback. In the first study, we investigated if feedback in the form of open-ended and thought-provoking questions can be crowdsourced and if these questions contain more neutral sentiment compared to corresponding feedback statements.
The results show that 85\% of the questions created by the crowd workers are open-ended and thought-provoking. We also found that the questions derived from negative or positive statements contained significantly more neutral sentiment than the corresponding statements, as exemplified in \autoref{fig:example-feedback}.
In the second study, we examined the effectiveness of feedback enhanced with open-ended questions on the quality of design revisions. We recruited 36 non-professional designers to design a flyer and revise it based on crowdsourced feedback. To test our hypothesis, we assessed three ways of presenting the feedback: statements only, questions only, and questions followed by statements.
\new{We employed an external jury of expert designers to rate the flyers' design quality for comparison.}
We found that participants who were shown questions followed by statements improved their designs to a significantly greater degree than participants who saw either statements or questions alone.

We make two contributions to the area of crowdsourced design feedback. First, we introduce the first method for framing crowdsourced design feedback as questions and combining them with traditional feedback statements.
Second, we provide empirical evidence that presenting crowdsourced feedback in the form of open-ended questions followed by statements improves the quality of design revisions compared to presenting feedback as either statements or questions alone.
Combining statement-based feedback with open-ended questions is complementary to other strategies for enhancing the effectiveness of design feedback. Therefore, our approach can easily be integrated into existing crowdsourced design feedback systems to increase the overall productive value of the feedback for design revisions.

%% file: main-2-related-work-final.tex
\section{Related Work} \label{sec:related-work}

\subsection{Background}

\new{Within the inherently iterative design process, feedback is essential to evaluate the design's current state and generate revision ideas~\cite{sadler1989formative,hattie2007power}. Design studios are a fundamental element in design education, where students receive feedback in various types of critique sessions~\cite{schon1985design}. These critique sessions consist of a work presentation by the student followed by an individual critique from the teacher (i.e., ``desk crit''), multi-layered critique by a jury, or open feedback from other students~\cite{uluoglu2000design}. Ideally these sessions result in a dialogue for finding a common ground between one's own design intentions and the received feedback. In the professional practice, designers are seeking such detailed feedback from peers. Overall, design critiques provide in-depth analyses and foster a deep understanding of the designer's work~\cite{dannels2008beyond,connor2015discussing}. However, while providing rich feedback, critiques can be infrequent, time-consuming, and resource-intensive.
Therefore, designers may require additional feedback in preparation for the more structured critique sessions. Peers and online communities can provide such additional feedback but it can be limited in quantity~\cite{marlow2014rookie,xu2014voyant,luther2015structuring}, biased~\cite{tohidi2006getting,xu2014voyant}, and superficial~\cite{willett2012strategies,xu2012you}. Crowdsourcing is an approach to overcome these limitations~\cite{luther2014crowdcrit,greenberg2015critiki,yen2016social} and provide almost expert-quality feedback when elicited and structured effectively~\cite{yuan2016almost}.}

\subsection{Sentiment and Valence} \label{sec:rl:sentiment}

Prior research on crowdsourced feedback systems found that the sentiment of feedback impacts its perceived usefulness.
For example, Yuan et al.~\cite{yuan2016almost} found that ``positively written and emotional critiques received higher average ratings''. Their findings provide evidence that valence and arousal are positively correlated with designers' ratings of feedback.
Similarly, Nguyen et al.~\cite{nguyen2017fruitful} studied feedback on writing tasks and found that positive tone in critical feedback leads to better work quality overall.
Krause et al.~\cite{krause2017critique} systematically investigated the perceived usefulness of feedback along various dimensions such as length, specificity, or complexity. They found that the perceived usefulness peaks for feedback with neutral to very mildly negative sentiment.
Wu et al.~\cite{wu2017bitter} build upon these findings and studied the effects of presenting feedback with varying sentiments in different orders.
They present empirical evidence that showing negative feedback at the end improved the feedback's perception.

However, in contrast to the perceived usefulness, Crain et al.~\cite{crain2017share} studied the long-term effects of different types of feedback on design iterations in a large meta-study on feedback collected from Reddit. They found that longer and less positive feedback is predictive of a higher number of design iterations. Although the study could only take publicly shared iterations into account, it highlights a disparity between the perceived usefulness and the actual effectiveness of feedback with diverging sentiment.

Sargeant et al.~\cite{sargeant2008understanding} studied the impact of positive and negative feedback on the recipient. They found that negative feedback can evoke negative feelings, especially when the feedback disagrees with the recipient's self-perception. In this case, the recipient perceives the feedback to be addressed against themselves rather than the task at hand. Wu et al.~\cite{wu2018soften} confirmed these findings and additionally showed that balancing the valence of feedback can mitigate the impact of negative feedback on its perceived usefulness.

We hypothesize that framing feedback as a question will alleviate sentiment. Subsequently, we hypothesize that showing feedback in the form of questions prior to the traditional statement-based feedback will increase the feedback's overall acceptability.

\subsection{Reflection} \label{sec:rl:reflection}

The ultimate goal of feedback is to help improve the critiqued work. In order to achieve this goal, feedback needs to facilitate new productive ideas. Beyond direct feedback, reflection is another popular tool~\cite{schon1984reflective} in the design community to generate ideas for design revisions. See Baumer et al.~\cite{baumer2014reviewing} for a review on how reflection can be leveraged in the design process as a whole. In regard to feedback, Caroline Brandt~\cite{brandt2008integrating} showed that feedback alone might not always be sufficient. She suggests that combining feedback with a reflection task is generally superior.

Yen et al.~\cite{yen2017listen} confirmed this hypothesis by showing that reflection alone can be as beneficial as crowdsourced feedback.
They implement a reflective activity where designers have to respond to three generic questions about their design.
In their study, the combination of reflection and feedback led to the best design quality overall.
Moreover, Sargeant et al.~\cite{sargeant2009reflection} found that facilitated reflection can alleviate the distress caused by negative feedback and enhance feedback acceptance.

In this work, we build upon these findings and hypothesize that feedback in the form of questions will act as a lightweight reflective activity that promotes useful ideas for design revisions. Moreover, we extend previous reflection approaches by preceding a negative feedback statement with an open-ended question related to the same aspect of the design to help designers to better cope with potential distress caused by the negative feedback.

\subsection{Facilitative Questioning} \label{sec:rl:questioning}

For questions to be effective, they need to facilitate reflection and promote critical thinking. For instance, in evaluating writing, Knoblauch and Brannon~\cite{knoblauch1984rhetorical} have established an approach called ``Facilitative Response'', which argues that the reviewer should adopt a ``facilitative posture''. Instead of directly telling the writer what to do, the reviewer should raise open-ended questions to encourage the writer to think about their ideas and expressions more fully. Facilitative responses do not need to come in the form of questions, but studies have found questions to be an effective implementation.

For example, Carnine et al.~\cite{carnine1982effects} found positive effects for facilitative questioning in combination with feedback in teaching children.
Berghmans et al.~\cite{berghmans2012facilitative} studied the benefits of facilitative questioning against direct teaching approaches for medical students. They found that facilitative questioning is beneficial for students with less expertise. Interestingly, they also discovered that questioning alone is not perceived well as students demand information after facilitative questions were raised.

In general, questioning has been studied as a tool for teaching. For example, Alison King developed a technique called ``reciprocal questioning''~\cite{king1992facilitating, king1990enhancing} in which she provides evidence that thought-provoking questions lead to a deep discussion about topics and encourage critical thinking~\cite{king1995inquiring}.
Ciardiello et al.~\cite{ciardiello1998did} discuss how to identify and generate divergent questions to promote literacy.
Chambers et al.~\cite{chambers2006effects} compared questioning as a teaching tool for swimmers and found that deliberately delaying extensive amounts of feedback and replacing it with insightful questions elicits better reflection and ultimately improves the swimmers' technique.

In our approach, we implement facilitative questioning as a tool to promote reflection and critical thinking.

\subsection{Framing \& Structuring Feedback} \label{sec:rl:structuring}

Irrespective of the feedback's sentiment and reflective nature, the way a system elicits \new{and structures} feedback from non-expert crowd workers can change the \new{feedback's focus and quality}. For example, Hicks et al.~\cite{hicks2016framing} investigate three different ways of framing feedback. They found that asking for numerical ratings of the design leads to more explanatory feedback of lower quality.

\new{Sadler describes effective feedback to be specific (following a predefined concept), goal-oriented (comparing the work's current to a reference state), and actionable (promoting actions that close the performance gap)~\cite{sadler1989formative}. As elaborated by Connor and Irizarry, these three elements are equally necessary for design critiques~\cite{connor2015discussing}. They additionally argued that the critique's goal should be an analysis of the performance gap to drive effective design iterations.
In the context of crowdsourcing,} several studies \cite{luther2015structuring,greenberg2015critiki,xu2014voyant,robb2015crowdsourced,yuan2016almost,ngoon2018interactive,kang2018paragon} have \new{evaluated} the effects of structuring and scaffolding feedback and found that an appropriate structure elicits more diverse and higher quality feedback.
For example, Voyant~\cite{xu2014voyant} prompts non-expert feedback providers to provide smaller feedback on various specific aspects of a design.
In CrowdCrit~\cite{luther2015structuring}, Luther et al. built upon these findings and further structured the feedback task into problem identification and explanation.

In our method, we utilize these findings by asking the feedback providers to focus on three different aspects of the design.

%% file: main-3-approach-hypotheses-final.tex
\section{Approach and Hypotheses} \label{sec:approch}

Previous research indicates a design tension (\autoref{sec:related-work}). Positive feedback is more acceptable to the recipient, but it is less likely to lead to substantial revisions compared to negative feedback. On the other hand, negative feedback can lead to substantial design improvements, but it is a source of discouragement and it is likely to be dismissed. This is particularly challenging in the context of crowdsourced design feedback systems, an otherwise promising source of feedback.
How can we enhance crowdsourced design feedback to be acceptable and substantive to promote useful ideas for design revisions? And how can we elicit such feedback robustly from non-expert crowd workers? 

Our approach is to structure feedback such that a potentially negative or positive statement is preceded by an open-ended question related to the same concern. For instance, in the context of designing an event flyer, ``This image is not relevant to the event'' might be preceded by ``What made you choose this image?'', or ``How is this image related to the event?''. To ensure that the question and statement relate to the same concern, the feedback provider is asked to first provide statement-based feedback and subsequently rephrase the statement into an open-ended and thought-provoking question. We consider a question to be open-ended when it requires an elaborating answer beyond ``yes'', ``no'', or simple facts. The goal of such a question is to promote critical thinking and reflection about a specific aspect of the critiqued work without carrying overly positive or negative sentiment.

\smallskip
In this context, our main hypothesis is the following:

\textbf{\heffect}: Feedback in the form of an open-ended question followed by a statement improves the overall quality of design revisions compared to statement-based or question-based feedback alone.
\new{Our reasoning is twofold.} We hypothesize that the preceding question increases the acceptance of negative feedback and that asking a question will act as a lightweight reflective task, which can promote better design revision, as shown by Yen et al.~\cite{yen2017listen}.
However, we expect feedback consisting of questions alone to lead to less effective design revisions as it can irritate the feedback receiver~\cite{berghmans2012facilitative}.

\smallskip
To answer our main hypothesis, we pose the following supporting hypotheses on the effects of question-based feedback:

\textbf{\hcrowd: Non-expert crowd workers can ask open-ended and thought-provoking questions.}
Given prior work on the effectiveness of structuring feedback acquisition (Section~\ref{sec:rl:structuring}), in particular the work by Greenberg et al.~\cite{greenberg2015critiki}, we hypothesize that providing a clear structure on how to provide feedback in combination with relevant example questions will teach the workers how to pose open-ended and thought-provoking questions, just like Alison King did with her students~\cite{king1992facilitating, king1990enhancing}.

\smallskip
\textbf{\hsenti: Feedback in the form of an open-ended question has more neutral sentiment than feedback addressing the same concern, but framed as a statement.}
Assuming that crowd workers are able to pose such questions, we hypothesize that open-ended questions carry more neutral sentiment than statements given the nature of open-ended questions. 

\smallskip
\textbf{\haccept: Preceding question-based feedback leads to more balanced acceptance of subsequent statement-based feedback compared to statement-based feedback alone.}
Assuming open-ended questions contain more neutral sentiment than statements and taking into account the improvement in perception of negative feedback when preceded by positive feedback~\cite{wu2017bitter}, we hypothesize that presenting the question-based feedback first will cause the recipients to focus on the design rather than themselves and perceive subsequent statement-based feedback more neutrally compared to statement-based feedback alone.

%% file: main-4-manipulation-check-final.tex
\section{Study 1: Eliciting Open-Ended Feedback Questions From Crowd Workers} \label{sec:manipulation-check}

In support of \heffect, we investigated if open-ended question-based feedback can be crowdsourced from non-experts (\hcrowd{}) and if such question-based feedback contains more neutral sentiment than statement-based feedback (\hsenti{}). To this end, we asked online crowd workers to provide feedback for graphic designs in the form of statements and questions.

\subsection{Experimental Design} \label{sec:manipulation-check:design}

In our approach (\autoref{sec:approch}), we ask each feedback provider to rephrase their feedback statement into a question to ensure that the feedback addresses the same aspect of the design. However, the act of rephrasing might be a confounding factor that influences the sentiment \new{and open-endedness}.
To control for this potential confounding factor, we conducted a within-subjects experiment with two factors: \emph{framing} and \emph{rephrasing}. Framing \new{has two levels, which} refer to posing feedback as either declaratory statements or open-ended questions. Rephrasing describes the strategy of eliciting statements-questions pairs \new{and has the following two levels}: rephrasing statements into questions (\sq{}) or vice versa (\qs{}).

\subsection{Task}
We presented each participant with four diverse designs of a flyer advertising a local event.
We asked each participant to provide three written feedback items (addressing the \textit{theme} of the design, the \textit{layout} of the design, and a specific visual \textit{element} in the flyer). For the first two flyers, the participants had to write a statement first and then rephrase it into a question (\sq{}). For the other two flyers, the participant had to first write the question and then rephrase the question into a statement (\qs{}).
Following Greenberg et al.~\cite{greenberg2015critiki}, we provided three diverse examples to promote creativity~\cite{siangliulue2015providing, siangliulue2015toward} and encourage feedback that addresses a variety of aspects. Each example consisted of a statement and question.

\subsection{Participants}
We recruited 24 participants (16 male and 8 female) on Amazon Mechanical Turk (AMT) who were located in the US and spoke English natively. Only participants with an acceptance rate above 97\% and more than 500 approved HITs were accepted. The majority of participants (16) were aged between 30--40. Three were between 20-30 years old. Another three were between 40-50 years old. And two were aged between 50--60. On average, the participants reported to be somewhat familiar with graphic design principles (M=3.17) and not very proficient in generating graphic designs (M=2.58). The results were reported on a 5-point Likert scale from ``very unfamiliar'' to ``very familiar'' and ``very unproficient'' to ``very proficient'' respectively. Participants were paid 5 USD for completing the task.

\subsection{Procedure}

We divided the participants into two groups, where the first group started with rephrasing statements into questions (\sq{}) two times and then switched to \qs{}. The second group started with \qs{} and switched to \sq{} after the first two flyers. Supplementary Figures S2--S4 show how the task was implemented. To avoid mistakes when the participants switched from \sq{} to \qs{} and vice versa, we added a dedicated step to inform about the upcoming switch in the rephrasing strategy. In total, each participant provided 12 feedback items: three feedback items for each of the four flyer designs. The order of the flyers was randomized. 

\subsection{Measurements} \label{sec:q-ana}

\paragraph{Open-endedness.} We measured the rate of successfully-rephrased statements into open-ended and thought-provoking questions through coding. The first two authors of this paper coded all statements as being either successfully rephrased into open-ended and thought-provoking questions or not. We considered a question to be open-ended and thought-provoking if it required more than a yes/no answer or a statement of simple facts. Specifically, we used Alison King's~\cite{king1990enhancing,king1992facilitating,king1995inquiring} question stems (e.g., ``How did you choose\ldots'', ``What is the purpose of\ldots'', or ``Why did you decide on\ldots'') as guidance and we assessed if the question targeted the rationale behind a design choice.

Prior to the analysis, feedback that did not target the actual design was removed. Such peripheral feedback questions typically focus on predefined requirements (e.g., ``What made you name it Harvard Open Boathouse if it's technically not "open" to anyone except for Harvard students?'') or facts about the photographic material (e.g., ``Is this one of the actual boats that are currently being used by the crew?''). 

The authors initially coded all questions individually \new{using separate Google Sheets with questions in randomized order. They achieved} high agreement of Krippendorff's $\alpha=.81$ \new{(calculated in \textit{Python} using Grill's \texttt{krippendorff\_alpha} method~\cite{grill2020})}. Subsequently, they collaboratively resolved conflicts to reach complete agreement. Most conflicts were due to two types of questions: questions that ask for a reason (e.g., ``Is there some reason why you did not decide to go with a more blue color to kind of go along with boating?'') and questions that ask for an alternative (e.g., ``Does the text at the bottom contrast enough against the water? Is there another color that might work better?'').

\paragraph{Sentiment.} We analyzed the sentiment of every feedback statement and question using \texttt{VADER}~\cite{hutto2014vader}---an automated sentiment analysis tool. \texttt{VADER} provides a polarity score ranging from $-1$ to $1$, where $-1$ refers to negative sentiment, $1$ refers to positive sentiment. We consider scores between $-0.05$ and $0.05$ as neutral sentiment.

\subsection{Results}
Ten out of 288 feedback questions ($3.5\%$) were removed from the analysis as they did not pertain to the graphical design choices.
Of the remaining 278 questions, 236 ($84.9\%$) were found to be open-ended and thought-provoking. 

\begin{figure}
  \centering
  \includegraphics[width=1\columnwidth]{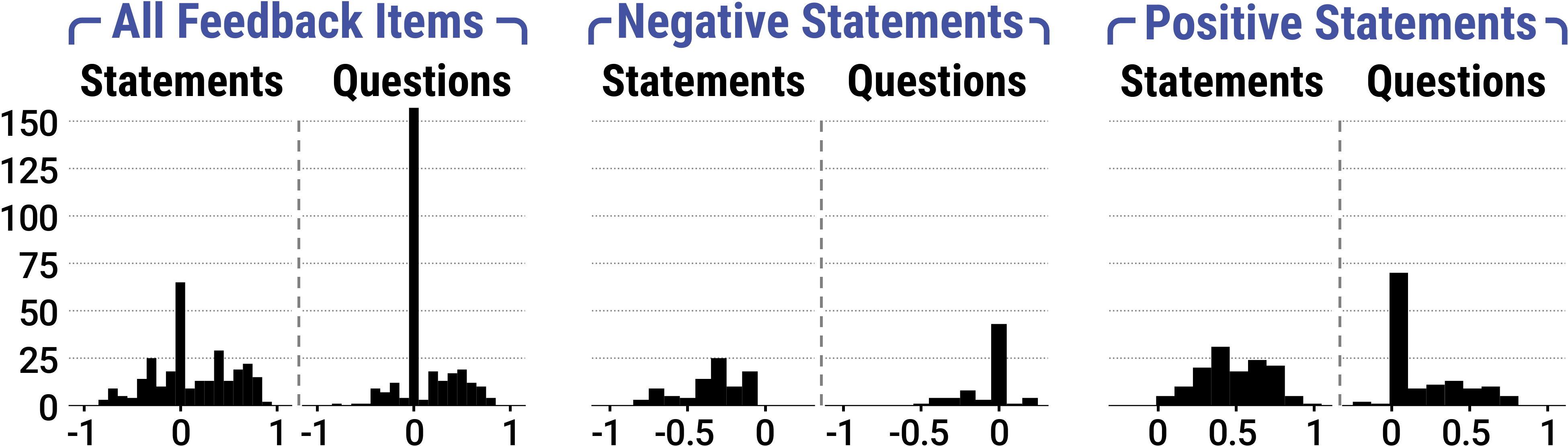}
  \Description[Three bar charts showing the distribution of feedback polarity]{Three bar charts showing the polarity distribution of statement-based and question-based feedback. Statements show stronger positive and negative polarity than questions. Both feedback types peak around neutral sentiment.}
  \caption{\textbf{Feedback Polarity:} \new{Distribution of} polarity scores \new{(x-axes)} \new{across} all feedback items (left), items related to negative statements (middle), and items related to positive statements (right). Questions have more neutral sentiment on average than the corresponding statements.}
  \label{fig:feedback-polarity}
\end{figure}

The distribution of \new{sentiment} polarity scores for the statement- and question-based feedback items are shown in~\autoref{fig:feedback-polarity}. \new{As confirmed by a Shapiro-Wilk test of normality, the polarity scores are not normally distributed (W=.92, p<.0001). Therefore, we conducted a Wilcoxon signed-rank test} to compare the absolute polarity of statement-based and question-based feedback. We found that statement-based feedback had significantly higher absolute polarity (M=.33, SD=.27) than question-based feedback (M=.18, SD=.23; \new{W=5703.5, p<.0001}). 

\new{To better understand how the question and statement sentiment differed}, we separately analyzed the polarity scores \new{of statement-question pairs associated to} statements with a polarity smaller than $-.05$ (i.e., negative statements), larger than $.05$ (i.e., positive statements), and polarity in [-.05, .05] (i.e., neutral statements). For negative statements \new{(n=87)}, we found that statement-based feedback had significantly more negative polarity scores (M=-.34, SD=.20) than the related question-based feedback (M=0.07, SD=0.28; \new{W=112}, p<.0001). Similarly, for positive statements \new{(n=128)}, statement-based feedback had significantly higher polarity scores (M=.50, SD=.21) than the related question-based feedback (M=.17, SD=.27; \new{W=643.0}, p<.0001). For neutral statements \new{(n=63)}, we did not find any significant difference in the scores for statement-based (M=.00, SD=.01) and question-based feedback (M=.04, SD=.21; \new{W=89.5, p=.14}).

\new{To determine the influence of \textit{rephrasing} (\sq{} and \qs{}), which might be a potential confounding factor (\autoref{sec:manipulation-check:design}), we analyzed its impact on the questions' open-endedness and sentiment. Knowing the influence of rephrasing can also inform future practical uses of our method.}
A \new{Cochran's Q test} showed that there was no significant association between rephrasing and open-endedness of the questions \new{(Q=8.92, p=.63)}.

\new{Regarding the impact of rephrasing on the sentiment polarity scores, we were additionally interesting in testing for potential interactions effects between rephrasing and framing. To use a nonparametric factorial analysis, we first applied the Aligned Rank Transform~\cite{wobbrock2011aligned} on the polarity scores. Using the aligned polarity scores,} we conducted a repeated-measures analysis of variance (ANOVA) with framing and rephrasing as the two within-subjects factors. As expected, we observed a significant effect of framing on absolute polarity (\new{F(1,552)=65.51, p<.0001}) and no significant effect of rephrasing on the absolute polarity (\new{F(1,552)=1.23, p=.27}). We also did not find any significant interaction between framing and rephrasing (\new{F(1,552)=1.46, p=.23}).

We separately repeated the same analysis for question-statement pairs associated with negative and positive statements. For negative statements, we again find a significant effect for framing (\new{F(1,170)=191.98, p<.0001}) and no significant effect for rephrasing (\new{F(1,170)=.53, p=.47}). \new{However, this time} we found a significant interaction between framing and rephrasing (\new{F(1,170)=5.41, p=.021}). \new{Investigating the simple main effects for \qs{} and \sq{} separately, we find that questions (M=.09, SD=.3) had a more neutral polarity score (\qs{}: M=.09, SD=.3; \sq{}: M=.05, SD=.27) than statements (\qs{}: M=-.38, SD=.21; \sq{}: M=-.31, SD=.19) in both cases (\qs{}: F(1,86)=110.87, p<.0001; \sq{}: F(1,84)=81.1, p<.0001).} Similarly, for positive statements, we find a significant effect for framing (\new{F(1,252)=115.92, p<.0001}), no significant effect for rephrasing (\new{F(1,252)=.96, p=.33}), and a significant interaction between framing and rephrasing (\new{F(1,252)=6.84, p=.01}) \new{We again investigated the simple main effects for \qs{} and \sq{} separately and found that questions had a more neutral polarity score (\qs{}: M=.19, SD=.25; \sq{}: M=.15, SD=.30) than statements (\qs{}: M=.47, SD=.21; \sq{}: M=.52, SD=.22) in both cases (\qs{}: F(1,128)=48.95, p<.0001; \sq{}: F(1,124)=68.67, p<.0001).}

\subsection{Summary and Discussion}
The results of this study demonstrate that non-experts recruited online can produce open-ended questions with a high degree of success \new{($84.9\%$)}, which supports \hcrowd{}.
Our results also demonstrate that feedback phrased as questions has weaker polarity than equivalent feedback presented as declarative statements according to automated sentiment analysis. That is, questions related to negative feedback express more neutral sentiment than their corresponding statements, and questions related to positive feedback also express more neutral sentiment than statements expressing equivalent critique. These findings support \hsenti{}.
Finally, our results suggest that \new{the order in which feedback is rephrased does not have a strong effect on the feedback's sentiment. While we see an interaction between framing and rephrasing, the simple main effects indicate that questions have significantly less sentiment compared to statement in both rephrasing orders.}

\new{One concern is the influence of the payment on the feedback. Prior research suggests that the principal effect of payment is the increased quantity of work: Unpaid crowds provide less feedback than paid workers~\cite{xu2012you,xu2014voyant}. A factor that may be of greater relevance is anonymity, which can improve the feedback quality by avoiding peer pressure~\cite{marlow2014rookie}. Thus, we assume that our results on the quality and sentiment of feedback will generalize to unpaid settings as long as the feedback is anonymous. However, more studies are necessary to verify this assumption.}

%% file: main-5-user-study-final.tex
\section{Study 2: The Effects of Combining Statement- With Question-Based Feedback} \label{sec:user-study}

In the second user study, we examined our main hypothesis \heffect{} and the supporting hypothesis \haccept{} in the context of a graphic design task. The study consisted of two sessions. In the first session, participants designed an event flyer, for which we subsequently crowdsourced feedback. Based on this feedback, participants revised their initial design in the second session. Finally, an independent jury of design experts rated the improvements of the revised designs.

\subsection{Experimental Design}

We conducted a between-subjects experiment in which we compared the following three conditions: statement-based feedback only (\cs{}), question-based feedback only (\cq{}), and question-based feedback followed by statement-based feedback (\cc{}). While our main hypothesis (\heffect{}) is that the revision quality in \cc{} will be higher than in \cs{}, we included \cq{} to be able to determine whether the hypothesized improvement is due to the combination or framing of feedback. The participants were equally and randomly distributed across the three conditions.

\begin{figure}[!b]
  \centering
  \includegraphics[width=0.95\columnwidth]{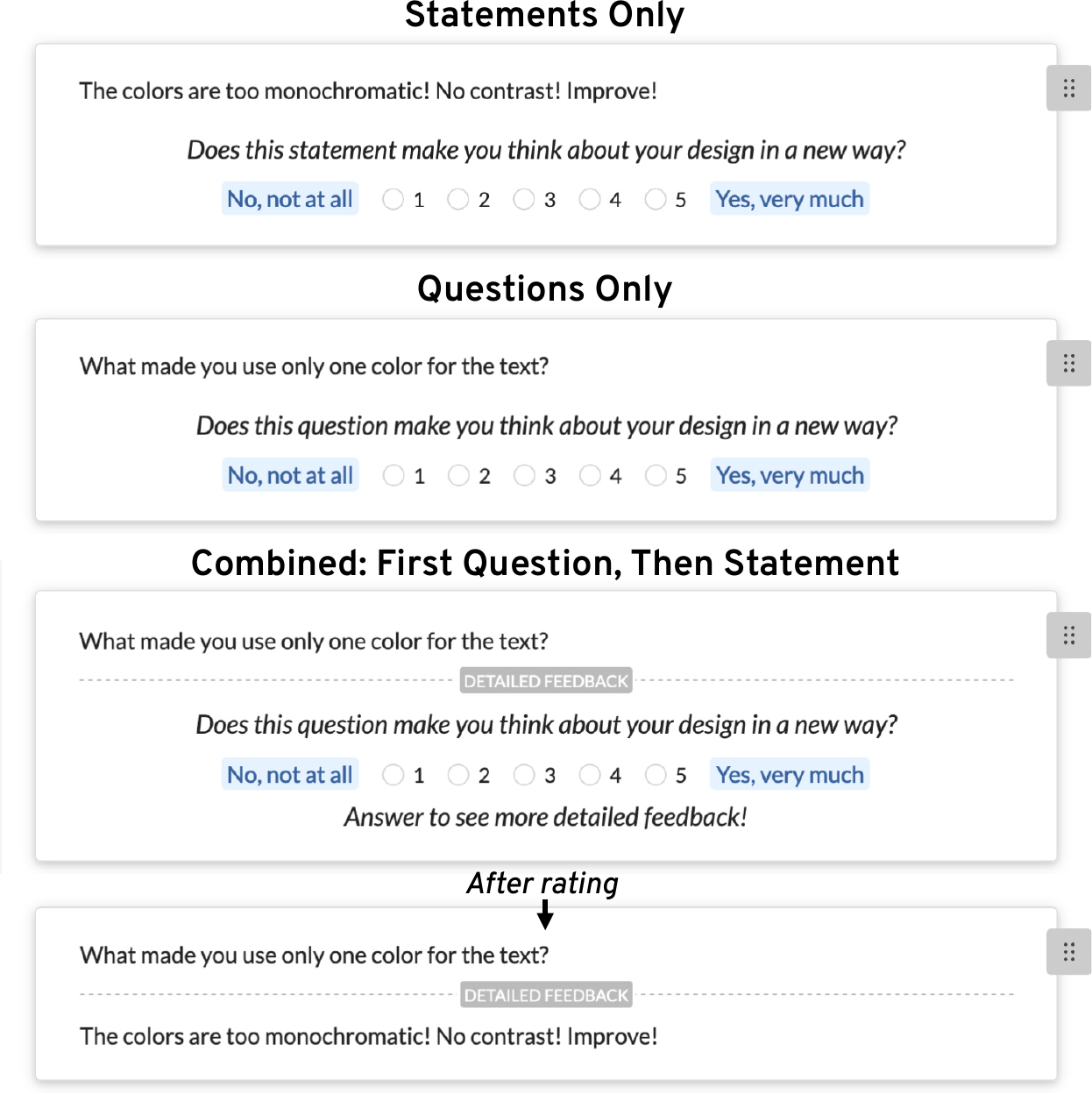}
  \Description[Screenshots of the feedback presentation and thought-provokingness rating.]{Four screenshots showing the user interface of the feedback presentation and thought-provokingness rating using a 5-point Likert scale.}
  \caption{\textbf{Feedback Presentation:} In the combined condition (\cc{}), the statement was only shown after the thought-provokingness was rated.}
  \label{fig:feedback-presentation}
\end{figure}

\subsection{Task} \label{sec:user-study:task}

The participants were asked to design a flyer for a local sports event. The event, called ``Harvard Open Boathouse''
was a fictional open house day of a university-affiliated rowing club that invites university members to learn about the sport, facilities, and meet senior club members. We chose this fictional event to focus on a specific event type that is popular in the local area.

In the first session, participants designed their initial flyer, which they subsequently in the second session. 
Before revising their flyer design, the participants were presented with crowdsourced feedback (\autoref{fig:feedback-presentation}), which we asked them to address in their revision. See Supplementary Figure S14 for a full example. During the feedback presentation, participants had to rate how much each statement or question made them think about their design in new ways.
Since our goal was to capture the immediately-perceived \emph{thought-provokingness} of each feedback item, the form fields disappeared after the corresponding feedback was rated. In the \cc{} condition, the participants saw only the question-based feedback until they rated the thought-provokingness, but a text label indicated that more information (i.e., the feedback statement) would appear after rating.
In all conditions, participants were not allowed to proceed and upload their revised design until all feedback items had been rated.
Inspired by Yen et al.~\cite{yen2017listen}, we wanted the participants to think about the question-based feedback explicitly to encourage reflection. Furthermore, in \cc{}, we wanted to contrast the reported thought-provokingness against the final feedback ratings (\autoref{sec:user-study:measurements}) to assess whether preceding questions increase the perceived usefulness of the feedback.

\subsection{Participants} \label{sec:user-study:participants}

\paragraph{Designers}
We recruited 36 participants (8 male and 28 female) located around Harvard University (Cambridge, MA) using flyers and mailing lists. The majority of participants (21) were aged between 18--25 while the rest (15) were aged between 26--35. We targeted participants who were relatively inexperienced in graphic design, as prior research~\cite{berghmans2012facilitative,dow2012prototyping} has shown that experienced designers have often built high confidence in their skill sets and rely primarily on their experience rather than feedback. 
In a pre-study questionnaire, most participants (25 out of 36) reported that they had never created a graphic design in a professional capacity. Per completion of both sessions, participants received a 35-USD gift card.

\paragraph{Feedback Providers}
We recruited 187 participants on AMT to provide feedback on the flyer designs. As in the first study (\autoref{sec:manipulation-check}), we only accepted US-based workers with an acceptance rate above 97\% and more than 500 approved HITs. To prevent any potential learning effects and ensure an equal distribution of independent feedback providers per design, we used Unique Turker~\cite{unique2020turker}, which stopped feedback providers from completing the user study multiple times. For statement- (\cs{}) and question-only (\cq{}) feedback, we paid 0.85 USD per task. For the combined feedback (\cc{}), we paid 1.25 USD per task.

\paragraph{Judges}
To evaluate and rate the improvement of the flyer designs, we recruited a jury of eight design experts \new{(three male and five female).} We considered someone to be a design expert if they hold an academic degree in a field related to graphic design, had at least two years of work experience as a professional designer or had taught at least one course related to graphic design. \new{Three experts earned a doctor degree while the others held a master degree in architecture, UI/UX/HCI, or fine arts. Five judges were professors, two were graduate research assistants with teaching experience, and one was a professional designer.} Each expert received a 50-USD gift card as compensation.

\begin{figure*}
  \centering
  \includegraphics[width=1\textwidth]{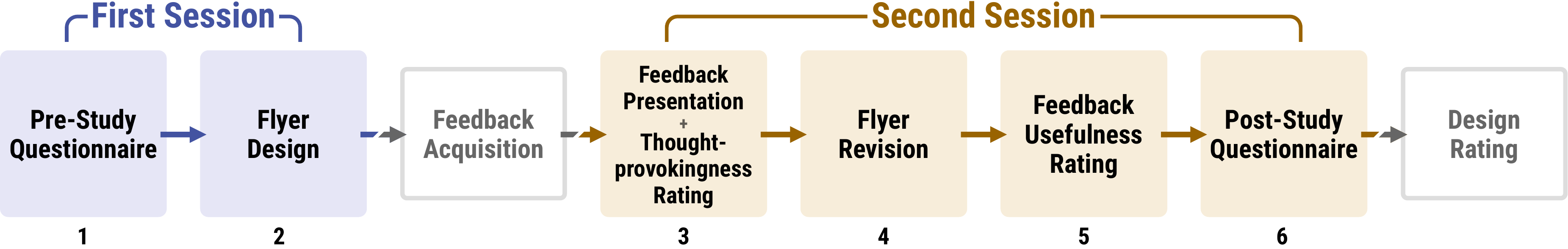}
  \Description[Flow chart of the user study procedure]{In the first session, the participants started by completing a pre-study questionnaire and then created an initial flyer design. Afterward, we crowdsourced feedback from AMT. In the second session, the participants first read the feedback, then revised their design, then rated the feedback, and finally completed a post-study questionnaire. At the end, a jury of design experts rated the improvement of the flyer designs.}
  \caption{\textbf{User Study Procedure:} In the first session, the participants completed a pre-study questionnaire (1) and created an initial flyer design (2). Afterward, we crowdsourced feedback from AMT. \new{(See ~\autoref{fig:example-feedback} for an example.)} In the second session, the participants read the feedback (3), revised their design (4), rated the feedback (5), and completed a post-study questionnaire (4). Finally, a jury of design experts rated the improvement of the flyer designs \new{(\autoref{fig:designs})}.}
  \label{fig:study-overview}
\end{figure*}

\subsection{Main Study Procedure} \label{sec:user-study:setup}

We conducted the study online to allow participants to work on their designs anywhere and anytime. Our web application guided the participants through each step of the user study. See Supplementary Figures S5--S19 for a complete walkthrough.
We split the experiment into two sessions to allow for enough time to collect feedback. \autoref{fig:study-overview} shows an overview of the procedure.

The first session comprised the consent process, pre-study questionnaire, design brief, and the first design iteration. The participants were free to use their software of choice for designing the flyer. For participants who did not have access to any graphics software, we recommended Google Drawings~\cite{google2020drawings} and Gravit Designer~\cite{gravit2020designer}.
After each participant completed the first session, we acquired, filtered, and randomly selected crowdsourced feedback. In the second session, the participants were presented with the feedback, revised their initial design, rated the received feedback, and completed the post-study questionnaire. Each session took 45--60 minutes. We started measuring the time before presenting the instructions for designing and revising the flyer and showed a timer for convenience.

Finally, an independent jury of design experts rated the improvement of the design revisions and selected the three best designs. We randomized the order of the flyers for each jury member to avoid interaction effects between the flyer's position and rating. The participant with the highest average quality rating received an additional 100-USD gift card. We included the competition to increase the participants' motivation throughout the two sessions.

\subsection{Acquisition and Selection of Crowdsourced Feedback} \label{sec:user-study:feedback-acquisition-selection}

For each flyer design, we collected 15 feedback items from five unique crowd-workers (i.e., three feedback items per worker) using the \sq{} feedback acquisition procedure from~\autoref{sec:manipulation-check}. Anticipating how the \cs{} and \cq{} conditions might be implemented in practice, we asked the feedback providers to only give statement-based or question-based feedback, respectively. Hence, the rephrasing step was omitted in \cs{} and \cq{}.

\begin{figure*}
  \centering
  \includegraphics[width=1\textwidth]{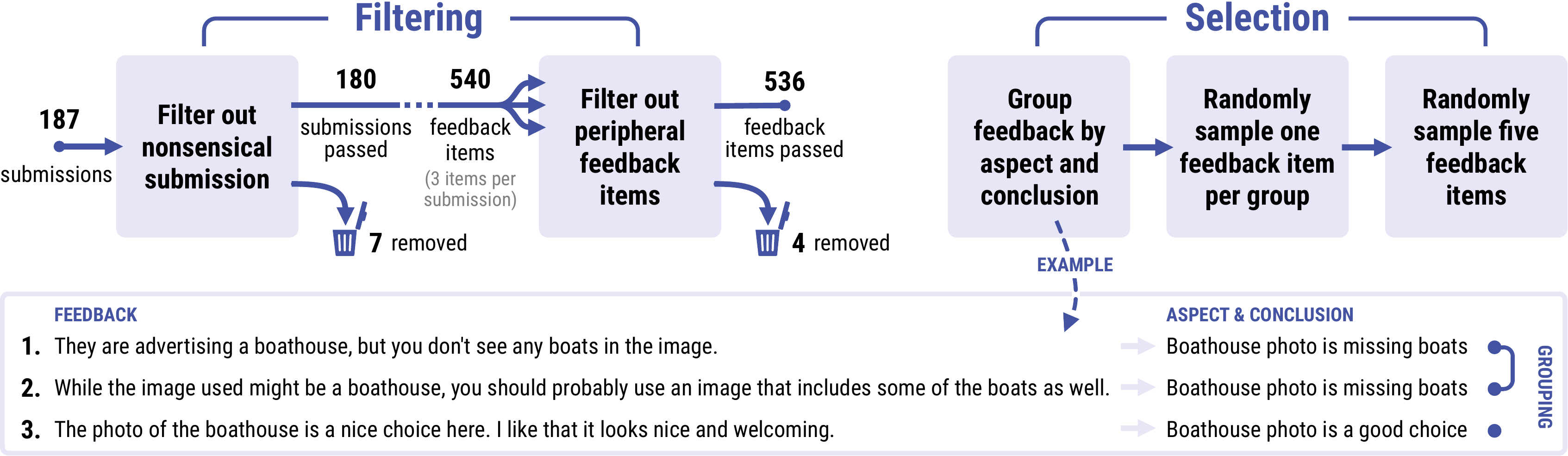}
  \Description[Flow chart of our feedback selection procedure]{First, we rejected nonsensical submission and removed peripheral feedback items. For each flyer design, we grouped the feedback by the main aspect (e.g., font size) and conclusion (e.g., too small) and randomly selected one feedback item per group. From the remaining feedback items we randomly sampled five feedback items that were presented to the participant.}
  \caption{\textbf{\new{Feedback Selection:}} First, we rejected nonsensical submissions and removed peripheral feedback items. \new{Next, f}or each flyer design, we grouped the feedback by the aspect (e.g., font size) and conclusion (e.g., too small) and randomly selected one feedback item per group. From the remaining feedback items we randomly sampled five feedback items that were presented to the participant.}
  \label{fig:feedback-selection}
\end{figure*}

After collecting the feedback (\autoref{fig:feedback-selection}), the first two authors of this paper inspected each set of three feedback items to ensure a minimum level of quality. In 7 out of 180 cases, the crowd worker provided incomprehensible or nonsensical answers (e.g., ``Element is Fine text''). We rejected these submissions and obtained new feedback. From the pool of 540 feedback items, we removed four peripheral feedback items that did not target the design itself, e.g., ``Why is the open boathouse restricted to only people with a university Harvard affiliations?''.

After filtering out invalid feedback, the first two authors of this paper grouped the feedback items that targeted the very same aspect of the flyer design and arrived at the same conclusion. For instance, as shown in~\autoref{fig:feedback-selection} (bottom), the three statements target the same visual element, but only the conclusion of the first and second are the same. Therefore, we grouped the first two but not the third feedback item. For each group, we randomly selected only one item. We used these groupings to avoid presenting the same critique multiple times. While the number of identical feedback items can provide an estimate for the critique's severity, we opted for diverse feedback instead.
Finally, we randomly selected five feedback items per design from the selection of unique feedback items, which were then shown to the participant during the second session. Given the time constraints for the revision task, we chose to limit the number of feedback items so that the participants did not have to spend much time on organizing the feedback.

\subsection{Measurements} \label{sec:user-study:measurements}

We used the results of three survey questions related to the feedback's thought-provokingness, usefulness, and tone as measures for the acceptance of feedback (\haccept{}). See Supplementary Figure S17 for an example.

\paragraph{Thought-provokingness.} In the second session, after having read each feedback item, but before submitting the revised design, we asked the participants: ``Does this [statement/question] make you think about your design in a new way?''. The participants provided their answers on a 5-point Likert scale ranging from ``no, not at all'' (1) to ``yes, very much'' (5).

\paragraph{Usefulness.} After the participants submitted their revised designs, we showed them the feedback again with the original and revised flyer design. This time, the participants had to rate each feedback item's usefulness in regards to the design revision by answering ``Was this feedback useful for revising your design?'' using a 5-point Likert scale ranging from ``no, not at all'' to ``yes, very much''. Our goal was to find out which feedback was perceived useful for revising the design as an indicator of the feedback acceptance.

\paragraph{Tone.} We also asked the participants to rate the tone of the feedback on a 5-point Likert scale from ``very negative'' to ``very positive'' to get a subjective rating of the feedback's sentiment polarity. To indicate that the tone is different from the feeling, we additionally asked the participants how the feedback made them feel.

\paragraph{Improvement.} To assess the impact of the feedback on the design revision (\heffect{}), we asked the jury members to rate the improvement of each flyer design on a diverging 7-point Likert scale ranging from ``worsened significantly'' (1) to ``significant improvement'' (7).

\subsection{Results}

The 36 participants created a total of 72 flyer designs (two designs per participant). \autoref{fig:designs} shows a diverse sample of eight flyer designs created by the participants.
The distributions of key measures per condition (\cs{}, \cq{}, and \cc{}) are shown in~\autoref{fig:feedback-ratings-improvement}.

To assess the overall effect of the feedback conditions on the quality of the design revisions, we analyzed the experts' improvement ratings of the revised flyers. The distribution is shown in~\autoref{fig:feedback-ratings-improvement} (right side). A \new{Kruskal-Wallis rank sum test} with condition (\cs{}, \cq{}, and \cc{}) as the independent variable and improvement as the dependent variable shows a significant effect of the conditions \new{(H=7.34, df=2, p=.0255)}. \new{A pairwise post-hoc Dunn test with Benjamini-Hochberg correction was significant for \cc{} versus \cs{} (p=.0341) and \cc{} versus \cq{} (p=.0479). However, \cs{} does not significantly differ from \cq{} (p=.89). The results} show that the mean improvement for \cc{} (M=4.77, SD=1.36) was significantly greater than the mean improvement for \cs{} (M=4.41, SD=1.14, d=.29) and \cq{} (M=4.32, SD=1.26, d=.34). The effect sizes for these analyses (d=.29 and d=.34) were found to exceed Cohen's~\cite{cohen1988statistical} convention for a small effect (d=.2).

A \new{Kruskal-Wallis rank sum test} with the condition (\cs{}, \cq{}, and \cc{}) as the independent variable and thought-provokingness as the dependent variable shows a significant effect of the condition \new{(H=10.17, df=2, p=.0061)}. \new{A pairwise post-hoc Dunn test with Benjamini-Hochberg correction was significant for \cs{} versus \cq{} (p=.0079) and \cc{} versus \cq{} (p=.0232). The results} show that the mean thought-provokingness of \cs{} (M=3.80, SD=1.22) and \cc{} (M=3.57, SD=1.25) were significantly higher than \cq{} (M=3.03, SD=1.33). However, \cc{} did not significantly differ from \cs{} \new{(p=.56)}.
Apart from that, we found no significant effect of condition on either usefulness \new{(H=3.62, df=2, p=.16)} or tone \new{(H=1.75, df=2, p=.42)}.

\new{To determine whether the feedback differed by some other measure, we conducted a Wilcoxon signed-rank test to compare the statement length between \cs{} and \cc{} and the question length between \cq{} between \cc{}. We found that statements in \cs{} (M=120.3, SD=52.4) are significantly longer than in \cc{} (M=87.8, SD=45.0; W=383.0, p<.0001). In contrast, the question length in \cq{} (M=90.2, SD=47.5) did not differ significantly \cc{} (M=90.9, SD=47.0; W=835.5, p=.88). We also compared the feedback's absolute polarity using a Wilcoxon signed-rank test but did not find any significant differences in the statements between \cs{} (M=.36, SD=.29) and \cq{} (M=.33, SD=.35; W=781.5, p=.56) and the questions in \cq{} (M=.15, SD=.21) and \cc{} (M=.2, SD=.27; W=341.5, p=.17).}

To verify if the redesigns were based primarily on the feedback obtained through this study, we asked participants after the study: ``Did you collect feedback or ideas for the revision elsewhere?'' (1 = ``no, not at all'' to 5 = ``yes, very much''). On average, the participants reported that they did not collect ideas elsewhere (M=1.39, SD=.99), and there was no significant difference between the conditions with respect to this question.

%% file: main-6-discussion-final.tex
\section{Overall Discussion}

\paragraph{Enhancing Feedback With Open-Ended Questions.}

In terms of the overall effect of \cs{} \new{(statements only)}, \cq{} \new{(questions only)}, and \cc{} \new{(question-based feedback followed by statement-based feedback)} on the quality of design revisions, we found that \cc led to significantly better revisions than either \cs{} or \cq{}, which provides evidence in support of our main hypothesis (\autoref{tab:results}). Even though the statement-based feedback we collected lacked strong sentiment on average, the effect sizes of \cc{} compared to \cs{} (d=.29) and \cc{} compared to \cq{} (d=.34) show a clear impact on the overall effectiveness of design feedback. \new{Such impact was not evident in previous work on enhancing crowdsourced design feedback~\cite{greenberg2015critiki,luther2015structuring,ngoon2018interactive}, which instead focused on improved feedback perception. The improvement in design iteration that we saw might in part be due to the reflective nature of question-based feedback. In this regard, our work extends the findings from Yen et al.~\cite{yen2017listen}, who demonstrated that a reflective activity alone can be as effective as feedback for design iterations. Yet, their results did not show a benefit of combining the reflective activity with traditional feedback, which was the case for \cc{} in our study.} Overall, we assume that the impact of \cc{} will be even greater in contexts where the crowdsourced feedback contains stronger sentiment, such as in social networks or web forums~\cite{yen2016social}.

\begin{figure}[b]
  \centering
  \includegraphics[width=1\columnwidth]{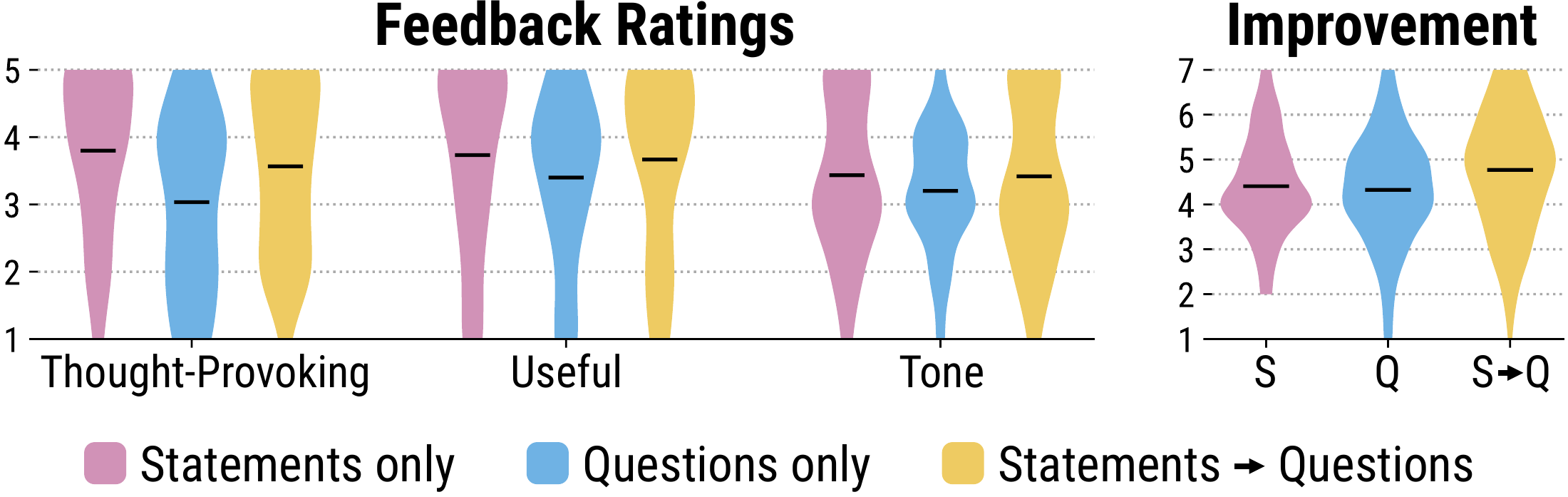}
  \Description[Four violin plots of the feedback rating and improvement measure distributions]{Violin plots showing the distribution of the thought-provokingness, usefulness, tone, and improvement measures split by condition. The usefulness and tone distributions are very similar. For the thought-provokingness, statements-only and statements+questions are greater than questions-only. Finally, the improvement in the statements+questions condition is visibly larger than statements-only and questions-only.}
  \caption{\textbf{Feedback Ratings and Design Improvements:} Distribution of the feedback ratings from the participants and improvement ratings of the jury. Note, the improvement score is provided on a diverging 7-point Likert scale where 1 refers to ``worsened significantly'' and 7 refers to ``significant improvement''.}
  \label{fig:feedback-ratings-improvement}
\end{figure}

\begin{figure*}
  \centering
  \includegraphics[width=1\textwidth]{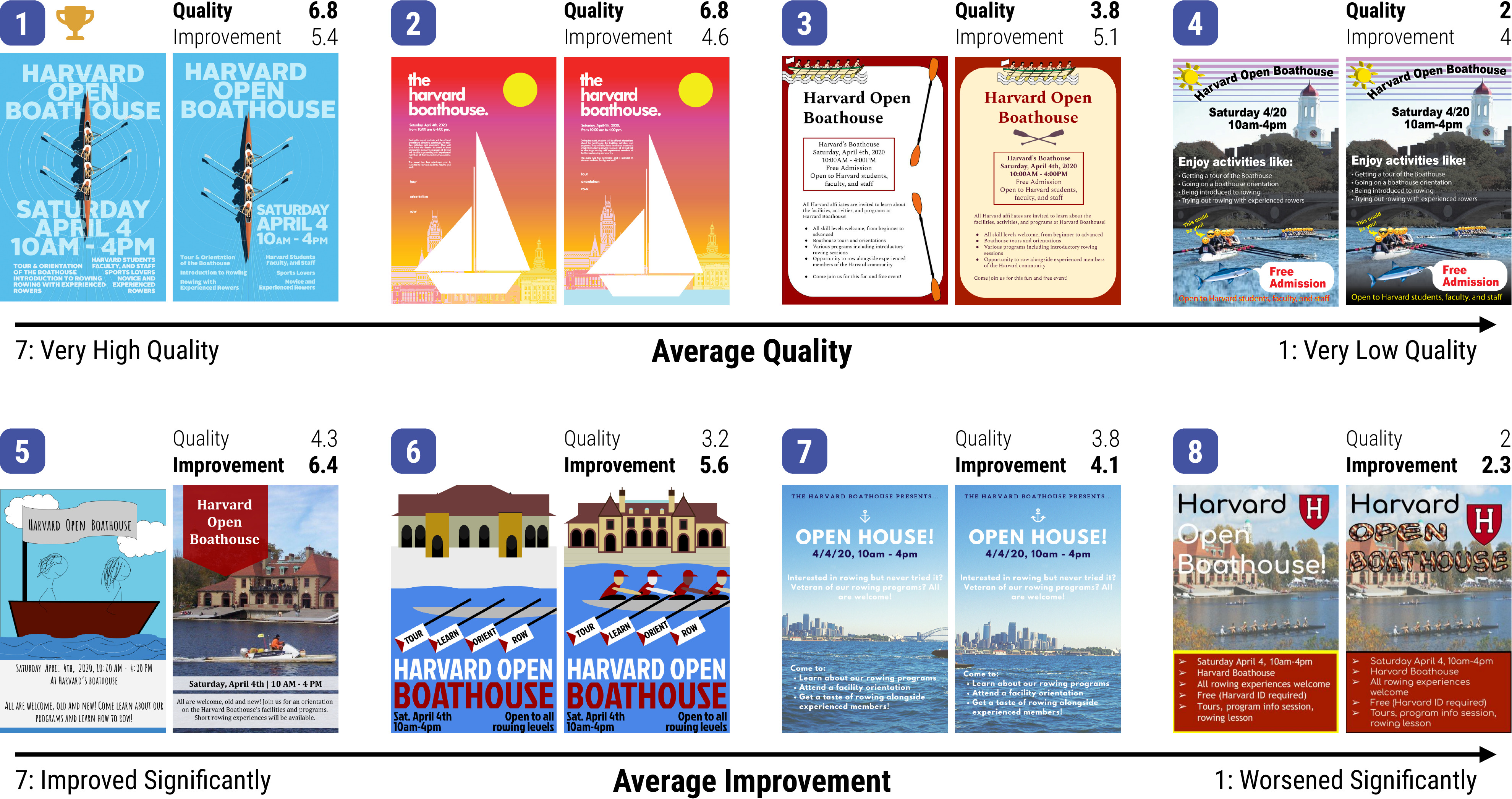}
  \Description[Eight flyer designs from our second user study]{Eight flyer design pairs (initial and revised design) with decreasing (left to right) quality (top row) and improvement (bottom row) scores.}
  \caption{\textbf{\new{Flyer Designs:}} Eight flyer designs \new{from study 2. The top row shows flyers} with decreasing \new{average quality scores of the revised design}. \new{The bottom row shows flyers with decreasing average improvement scores}. Each pair of images shows the original design on the left and the revised design on the right. The first flyer (1) won the best design award.}
  \label{fig:designs}
\end{figure*}

Furthermore, as expected, we found that feedback in the form of questions only (\cq{}) led to the least-improved design revisions. These results, albeit the difference between \cs{} and \cq{} was not significant, are in line with previous work~\cite{berghmans2012facilitative} and suggest that question-only feedback should not replace statement-based feedback for novices.

In support of our approach, through manually coding questions as either open-ended and thought-provoking or not, we show that it is indeed possible to enable online crowd workers to rephrase their statements into open-ended and thought-provoking questions. In total, 85\% of all questions were successfully rephrased, which we believe is a strong indicator that our AMT task design is an effective approach to crowdsource question-based feedback. Therefore, \hcrowd{} is supported. To further improve the success rate, future work could guide the elicitation of question-based feedback with natural language processing towards open-endedness.

The results of the polarity analysis strongly indicate that questioning is an effective technique to neutralize sentiment. In particular, the sentiment of negative statements is resolved entirely, which is essential to avoid negatively influencing the recipient's affective state. Interestingly, the sentiment of positive statements is also reduced, which suggests that question-based feedback carries less sentiment overall. In conclusion, our results suggest that \hsenti{} is supported. \new{By presenting question-based feedback prior to statement-based feedback, our method is an implementation of Wu et al.'s approach for mitigating unwanted effects of negative sentiment~\cite{wu2017bitter}.}

\begin{table}[b]
\begin{tabularx}{\columnwidth}{lXc}
  \toprule
  \small \textbf{Hypothesis} & &  \small \textbf{Support} \\
  \midrule
  \small \heffect & \small Feedback presented as questions followed by statements improves design revisions compared to statement-based or question-based feedback alone. & \small \textcolor{OkabeItoBlue}{\includegraphics[width=0.8em]{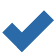}~Yes} \\
  \small \hcrowd & \small Non-expert crowd workers can ask open-ended and thought-provoking feedback questions. & \small \textcolor{OkabeItoBlue}{\includegraphics[width=0.8em]{main-figures-final/check.pdf}~Yes} \\ 
  \small \hsenti & \small Question-based feedback has more neutral sentiment than statement-based feedback. & \small \textcolor{OkabeItoBlue}{\includegraphics[width=0.8em]{main-figures-final/check.pdf}~Yes} \\
  \small \haccept & \small Feedback presented as questions followed by statements leads to more balanced acceptance of subsequent statement-based feedback. & \small \textcolor{OkabeItoPink}{\includegraphics[width=0.8em]{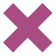}~No} \\
  \bottomrule
\end{tabularx}
\caption{\textbf{Key Findings:} The results support our main hypothesis and two out of three supporting hypotheses.}
\Description[Key Findings]{\new{The results support} our main hypothesis and two out of three supporting hypotheses.}
\label{tab:results}
\end{table}

Regarding the effects of questions on the perception of statements with overly positive or negative sentiment, we did not find any significant differences between the conditions in the reported usefulness ratings. Therefore, we cannot confirm \haccept{}. \new{In comparison, related work~\cite{greenberg2015critiki,luther2015structuring,ngoon2018interactive} found that structuring and scaffolding can improve the feedback's perceived usefulness. A potential explanation why we still saw an improved effectiveness of the \cc{} feedback compared to \cs{} and \cq{} could be that preposed question-based feedback primarily changes the recipient's focus from themselves to the design task. This change might have mitigated the effects of negative feedback~\cite{sargeant2008understanding}}. Contrary to our expectations, the only significantly different feedback rating was thought-provokingness, which was the lowest in \cq{}. In hindsight, asking participants about the magnitude of how much a feedback item made them think about their design might have been too unspecific. For instance, instructional feedback could have prompted the participants to think a lot about how to execute suggestions rather than to think about alternative designs. A more in-depth analysis of the revised designs could uncover which feedback was indeed addressed. It might also be necessary to study this question by limiting the feedback to highly negative and positive statements to emphasize the potential effect of questions on the perceived usefulness.

\paragraph{Generalizability.}
Given the breadth of related work, we would assume to see similar effects of question-based feedback in other domains. In particular, question-based feedback should easily be applicable to different areas of creative work due to the similar processes of iteration. Regarding our method for crowdsourcing question-based feedback, there are no technical limitations to expanding this method to other types of work. However, the success of crowdsourcing question-based feedback depends on the accessibility of the work to non-expert crowd workers. While graphic design in general and flyer-based advertisement in specific should be accessible by most people, this might not be the case for other types of work.

Beyond crowdsourcing, questions could also be employed as a generic method to enhance feedback. However, the usefulness of question-based feedback might be limited by the ability of the feedback providers to ask effective questions. More work needs to be done to better understand how the effectiveness of questions and statements are related when the feedback is obtained in other contexts, for instance, from domain experts.

\paragraph{Limitations.}
On average, the design revision improvement across all conditions was in line with previous work on the effectiveness of crowdsourced feedback~\cite{luther2015structuring}. However, by splitting the second study into two separate sessions, we might have lowered the participants' motivation and excitement, as they were compensated only after completing both sessions. An effort-based compensation approach might have helped to increase the participants' motivation.

\new{In this study we focused on the feedback's effectiveness for design iteration. In terms of the perceived feedback quality, we did not find any differences except for the thought-provokingness. And while the statement lengths differed between \cs{} and \cc{}, it is unclear how to interpret the comparison given that \cc{} additionally included the  questions. One option to generically quantify the quality could be to ask designers to enumerate revision ideas prior to the actual redesign, which we leave as an idea for future work.}

More fundamentally, assuming that the statements and questions are of the same quality, questions can reduce the sentiment of feedback statements and potentially facilitate reflection, but they cannot make the feedback, as a whole, more substantive.

%% file: main-7-conclusion-final.tex
\section{Conclusion and Future Work}

In this study, we empirically compared the effectiveness of crowdsourced design feedback on design revisions when presented as statements, questions, and a combination of both. Our results show that the combination of question- and statement-based feedback leads to better design revisions. We believe that these findings are generalizable to other kinds of creative work beyond graphic design. Also, we regard presenting feedback as open-ended questions to be complementary to other approaches for improving crowdsourced feedback. Therefore, it can be integrated into existing online feedback systems to improve the overall effectiveness of crowdsourced feedback further.

Future studies may analyze how exactly questions influence the perception of related statements by exclusively examining feedback that carries strongly positive and negative sentiment, or explicitly letting the designer elaborate on their revision to relate changes to specific feedback items.
Moreover, it would be interesting to evaluate what aspects determine the quality of question-based feedback regarding reflection. We assume that, similar to statements, the ability of questions to generate productive ideas for design revisions depends on their specificity. However, more aspects likely come into play.
Also, given that designers with varying expertise make sense of and provide feedback differently~\cite{foong2017novice,dannels2008critiquing}, it would be interesting to determine if question-based feedback is perceived differently by non-professional and professional designers.